\documentclass{pasj00}

\begin{document}
\SetRunningHead{Matsushita \& Chen}
	{Testing the ACA Phase Correction Scheme}
\Received{2010/02/23}%{yyyy/mm/dd}
\Accepted{2010/05/17}%{yyyy/mm/dd}
%\Published{}%{yyyy/mm/dd}

\title{Testing the ACA Phase Correction Scheme using the SMA}

\author{Satoki \textsc{Matsushita}\altaffilmark{1,2}
	and Yu-Lin {\sc Chen}\altaffilmark{1,3,4}}
\altaffiltext{1}{Institute of Astronomy and Astrophysics,
	Academia Sinica, P.O.~Box 23-141, Taipei 10617, Taiwan, R.O.C.}
\email{satoki@asiaa.sinica.edu.tw}
\altaffiltext{2}{Joint ALMA Office, Av. El Golf 40, Piso 18,
	Las Condes, Santiago, Chile}
\altaffiltext{3}{Department of Physics, National Taiwan University,
	No.~1, Sec.~4, Roosevelt Road, Taipei 10617, Taiwan, R.O.C.}
\altaffiltext{4}{Taipei County Yong-Ho Junior High School, No.~111,
	Guozhong Rd., Yonghe City, Taipei County 234, Taiwan, R.O.C.}


\maketitle

\begin{abstract}
We conducted the observational tests of a phase correction scheme for
the Atacama Compact Array (ACA) of the Atacama Large Millimeter and
submillimeter Array (ALMA) using the Submillimeter Array (SMA).
Interferometers at millimeter- and submillimeter-wave are highly
affected by the refraction induced by water vapor in the troposphere,
which results as phase fluctuations.
The ACA is planning to compensate the atmospheric phase fluctuations
using the phase information of the outermost antennas with
interpolating to the inner antennas by creating a phase screen.
The interpolation and extrapolation phase correction schemes using
phase screens are tested with the SMA to study how effective these
schemes are.
We produce a plane of a wavefront (phase screen) from the phase
information of three antennas for each integration, and this phase
screen is used for the interpolation and extrapolation of the phases
of inner and outer antennas, respectively.
The interpolation scheme obtains apparently improved results,
suggesting that the ACA phase correction scheme will work well.
On the other hand, the extrapolation scheme often does not improve
the results.
After the extrapolation, unexpectedly large phase fluctuations show
up to the antennas at the distance of $\sim140$~m away from the
center of the three reference antennas.
These direction vectors are almost perpendicular to the wind
direction, suggesting that the phase fluctuations can be well
explained by the frozen phase screen.
\end{abstract}

\section{Introduction}
\label{sect-intro}

The Atacama Large Millimeter and submillimeter Array (ALMA), the
largest millimeter and submillimeter interferometer ever built, is
currently under construction in the northern Chile with the
collaboration between East Asia, Europe, and North America
\citep{woo09}.
The ALMA is composed of up to eighty high-precision antennas, located
on the Chajnantor plain of the Chilean Andes in the District of San
Pedro de Atacama, 5000~m above sea level.
The ALMA covers the wavelength range from 0.3~mm to 9~mm with an
angular resolution of up to \timeform{0''.004}.
The Atacama Compact Array (ACA) is designed to improve the short
baseline coverage of the ALMA, especially for the observations of
extended structures at submillimeter wavelength \citep{igu09}.
The ACA consists of four 12~m telescopes to obtain the single-dish
data and twelve 7~m telescopes to obtain short baseline
interferometric data.

However, ground-base astronomical observations in the millimeter and
submillimeter ranges are strongly affected by the fluctuation of the
tropospheric water vapor distribution (e.g., \cite{tho01}).
Therefore, the correction of phase fluctuations due to the spatial
and temporal variations of the tropospheric water vapor content is
extremely important in millimeter and submillimeter interferometry.
Several kinds of techniques have been proposed and performed for
reducing phase fluctuations in millimeter and submillimeter
interferometry (see \cite{car99} for a review), such as fast
switching phase calibration \citep{hol92,hol95a,car97,mor00}, paired
array phase calibration \citep{hol92,car96,asa96,asa98}, and
radiometric phase calibration
\citep{lay97,car98,car99,mar98,del00,wie01,mat02}.
Although the ALMA site is one of the best sites for the millimeter
and submillimeter astronomy \citep{mat98,mat99,pai00,mat03},
atmosphere in this area is often affected by the phase fluctuations
due to water vapor \citep{rad96,hol97,but01}.
The phase correction methods for the ALMA and the ACA are therefore
highly needed to be considered and tested.

\citet{asa05} conducted a series of simulations of a phase
correction scheme for the ACA using water vapor radiometers (WVRs).
The WVRs can measure the tropospheric water vapor content directly
with observing a water vapor line in centimeter or millimeter
wavelength.
In the proposed ACA phase calibration scheme, the WVRs are attached
to the 12~m antennas at the four corners of the twelve 7~m array
(hereafter we call this as the reference rectangle).
The changes of the tropospheric water vapor content aloft measured
with the WVRs is transferred into the excess path lengths of the
arriving radio waves.
The excess path lengths measured at the four corners of the
reference rectangle are then fitted to a simple two-dimensional
slope or a screen.
Then the phases of the antennas inside the reference rectangle can be
compensated and calibrated.
Their simulation succeeded to compensate the atmospheric phase well,
which strongly support the use of the proposed phase correction
scheme.

To confirm this simulation study observationally and discuss further,
we performed the proposed phase correction scheme for the ACA using
the Submillimeter Array (SMA; \cite{ho04}).
Here we present measurements with the SMA at 230~GHz, analyze the
datasets under the proposed scheme, and discuss the results of the
corrected phase variations.
Our experiment is to clarify how effectively the proposed
compensation scheme works under the conditions of the real
atmosphere.
We construct a reference triangle composed of three antennas, and
make a flat phase plane or a screen with observing a strong point
source (section~\ref{sect-mdr}).
The phases of antennas inside the reference triangle can be
interpolated, while the phases of antennas outside can be
extrapolated.
We then compare the observed and the predicted phases of the point
source.
Standard deviations and temporal structure functions of the
observed (uncorrected) and the corrected phases are also compared
(section~\ref{sect-res}).
Finally, we discuss the usefulness of the interpolation and
extrapolation phase correction schemes using a phase screen, and
also discuss the validity of the ``frozen-flow'' hypothesis
(section~\ref{sect-dis}).

\section{Measurements and Data Reduction}
\label{sect-mdr}

The purpose of our experiment is to investigate observationally the
proposed phase compensation method for the ACA using the SMA.
The reasons of using the SMA for this experiment are that the SMA is
the interferometer operating at submillimeter wavelength, which is
the same as the ACA, and that the antenna configuration of the array
is applicable for this experiment.
The ACA uses the WVRs for the phase compensation, but for this
experiment using the SMA, since the SMA does not have WVRs, we
observed a strong point source to measure the phase directly to
perform the proposed ACA phase conpensation method.

\subsection{Measurements}
\label{sect-meas}

\begin{figure}
  \begin{center}
    \FigureFile(80mm,100mm){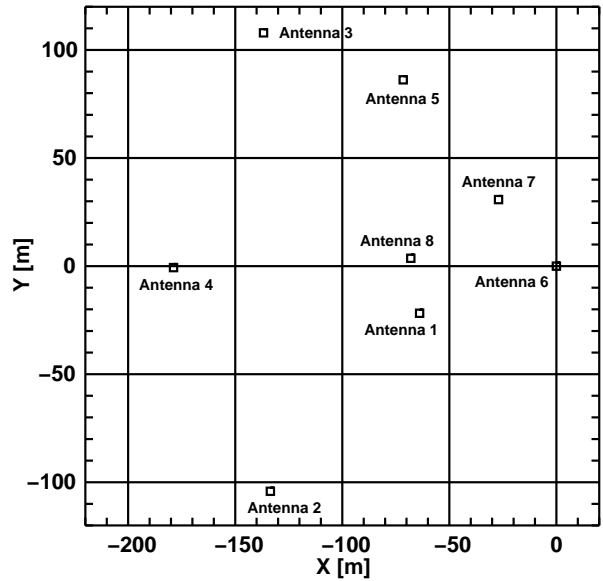}
  \end{center}
  \caption{Configuration of all eight antennas of the SMA at the
	observations.
  }
  \label{fig-conf}
\end{figure}

The measurements were carried out on August 26, 2004 using all eight
antennas of the SMA on Mauna Kea in Hawaii, and on September 7, 2004
using seven antennas (the antenna 8 was not used in this
measurement).
The antenna configuration is depicted in figure~\ref{fig-conf}.
The shortest and the longest baseline lengths are 11.6~m (antenna 1
to antenna 8) and 179.2~m (antenna 4 to antenna 6), respectively.
The measurements were performed at 240.0~GHz for the August 26, 2004
measurement and 230.5~GHz for the September 7, 2004 measurement
(both frequencies are at the upper side band).
Correlator bandwidth in both measurements was 2~GHz for each side
band.
We observed B$1921-293$ (J$1924-291$) in both days for 1.157 and
0.469 hours, respectively, with the integration time for one data
point (one integration number) of 5.16 seconds.
The integration number is the number of our data recorded every
integration time since the measurements started, and the ranges of
the integration number for the target source turned to be 2321 --
3128 and 140 -- 467 for the August 26 and September 7 measurements,
respectively.
The 230~GHz flux density of this source at the time was about 6 Jy,
strong enough to detect in one integration with high signal-to-noise
ratio.
The data were stored in the SMA data archive directories
040826\_01:23:23 and 040907\_05:22:09.
Hereafter we call the former dataset as ``040826'' and the latter
``040907''.

\subsection{Data Reduction}
\label{sect-data}

\subsubsection{Calibration}
\label{sect-data-calib}

We reduced the data using the Owens Valley Radio Observatory software
package MIR adopted for the SMA.
We only used the upper side band data.
The data were calibrated based on the antenna-base calibration.
We confirmed that all the final results did not change with the
reference antennas in the antenna-base calibration.

\subsubsection{Construction of a Phase Screen}
\label{sect-data-const}

\begin{figure}
  \begin{center}
    \FigureFile(80mm,100mm){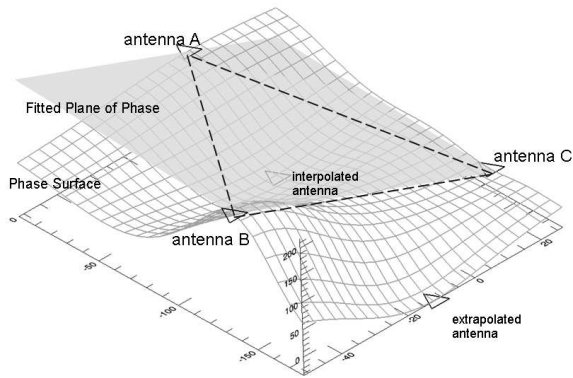}
  \end{center}
  \caption{Schematic diagram of a reference triangle with three
	antennas (A, B, and C) and a constructed phase screen.
    Interpolated and extrapolated antennas are also shown.}
  \label{fig-screen}
\end{figure}

Here we explain the procedure to construct a phase screen using a
reference triangle and the comparison between the observed and
estimated phases.
The schematic diagram is depicted in figure~\ref{fig-screen}.

\begin{table*}
  \begin{center}
  \caption{Configurations of reference triangles and interpolated and
	extrapolated antennas inside and outside the triangles.}
  \label{tab-conf}
  \begin{tabular}{cllll}
    \hline
    & \multicolumn{4}{c}{Datasets} \\
    \cline{2-5}
    & \multicolumn{2}{c}{040826} & \multicolumn{2}{c}{040907} \\
    \hline
     Reference  & Interpolated & Extrapolated  & Interpolated & Extrapolated \\
     Antennas   &  Antenna(s)  &   Antennas    &  Antenna(s)  &   Antennas   \\
    \hline
    $[2, 3, 6]$ &    1, 8      &    4, 5, 7    &      1       &    4, 5, 7   \\
    $[2, 4, 5]$ &    ---       & 1, 3, 6, 7, 8 &     ---      &  1, 3, 6, 7  \\
    $[2, 4, 6]$ &      1       &  3, 5, 7, 8   &      1       &    3, 5, 7   \\
    $[2, 4, 7]$ &	   8       &  1, 3, 5, 6   &     ---      &  1, 3, 5, 6  \\
    $[2, 5, 6]$ &   1, 7, 8    &     3, 4      &     1, 7     &      3, 4    \\
    $[3, 4, 6]$ &      8       &  1, 2, 5, 7   &     ---      &  1, 2, 5, 7  \\
    $[4, 5, 6]$ &    7, 8      &    1, 2, 3    &      7       &    1, 2, 3   \\
  \hline
  \end{tabular}
  \end{center}
\end{table*}

We first consider a reference triangle composed of three antennas
located at the outer periphery of the antenna configuration (antennas
A, B, and C in figure~\ref{fig-screen}).
Then, a plane of phase or a phase screen through these three antennas
can be constructed and computed at each integration (data point).
Note that since the geometrical delay is taken into account at the
correlation process (at the SMA backend), we do not need to consider
the effect of the antenna altitude differences.
In addition, we consider an antenna-base large timescale ($>20$
minutes) phase drift as an instrumental phase drift, so we subtract
this phase drift with second order polynomial fitting from the data
(this kind of large timescale phase drift will be taken out by the
phase calibration anyway in the real astronomical observations).
From this phase screen, interpolations and extrapolations are
conducted to predict phases at each antenna position inside and
outside the reference triangle, respectively.
We then compare the observed phases with our interpolated and
extrapolated phases.

We performed seven different configurations of reference triangles to
interpolate and extrapolate the phases of antennas inside and outside
the triangle, respectively.
The configurations of all the reference triangles we calculated are
summarized in table~\ref{tab-conf}.
For example, the reference antennas [2, 3, 6] means that we construct
phase screens with the coordinates and the observed phases of three
antennas 2, 3, and 6.

\section{Results}
\label{sect-res}

\subsection{Comparisons between Observed, Interpolated, Extrapolated,
	and Residual Phase Fluctuations}
\label{sect-res-comp1}

\begin{figure*}
  \begin{center}
    \FigureFile(80mm,100mm){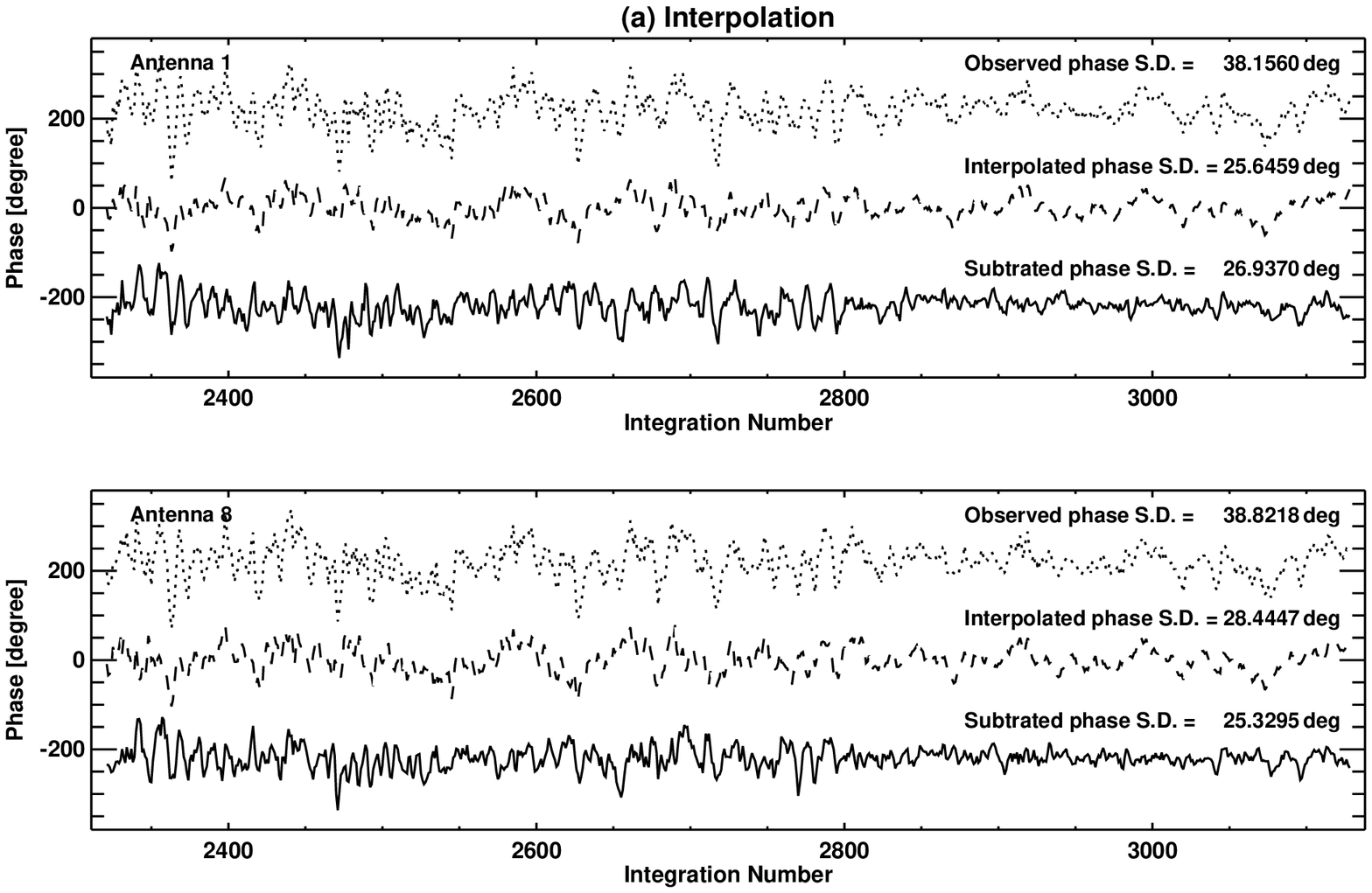}
    \FigureFile(80mm,100mm){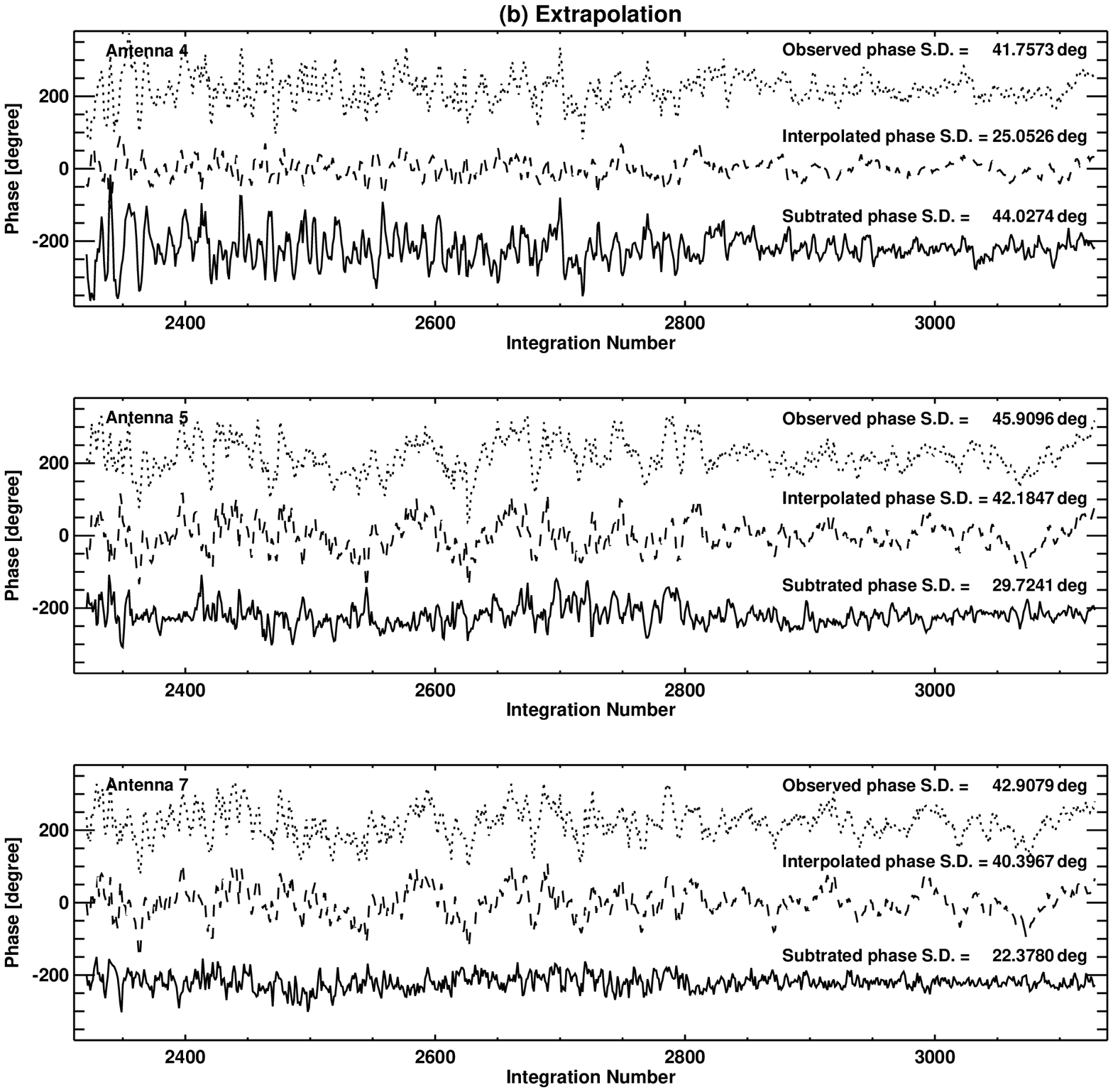}
  \end{center}
  \caption{The time series plots of various phases from the dataset
    040826 between the integration number 2321 -- 3128 with the
    reference triangle [2, 3, 6].
    The dotted curves show the observed antenna-base gain calibrated
    data, the dashed curves show the interpolated or extrapolated
	data, and the solid curves trace the subtracted data between the
    observed and modeled (interpolated/extrapolated) phases, namely
    the residual (phase corrected) data.
	The observed and subtracted data curves are arbitrary offset not
	to overlap each other.
    The standard deviation (S.D.) of the phase fluctuations over the
    integration time range for each curve are shown in the plots.
    The interpolated results always have better results, while the
    extrapolated results do not always have in this example.
    (a) The interpolation results for the antennas 1 and 8.
    (b) The extrapolation results for the antennas 4, 5, and 7.}
  \label{fig-resid1}
\end{figure*}

\begin{figure*}
  \begin{center}
    \FigureFile(80mm,100mm){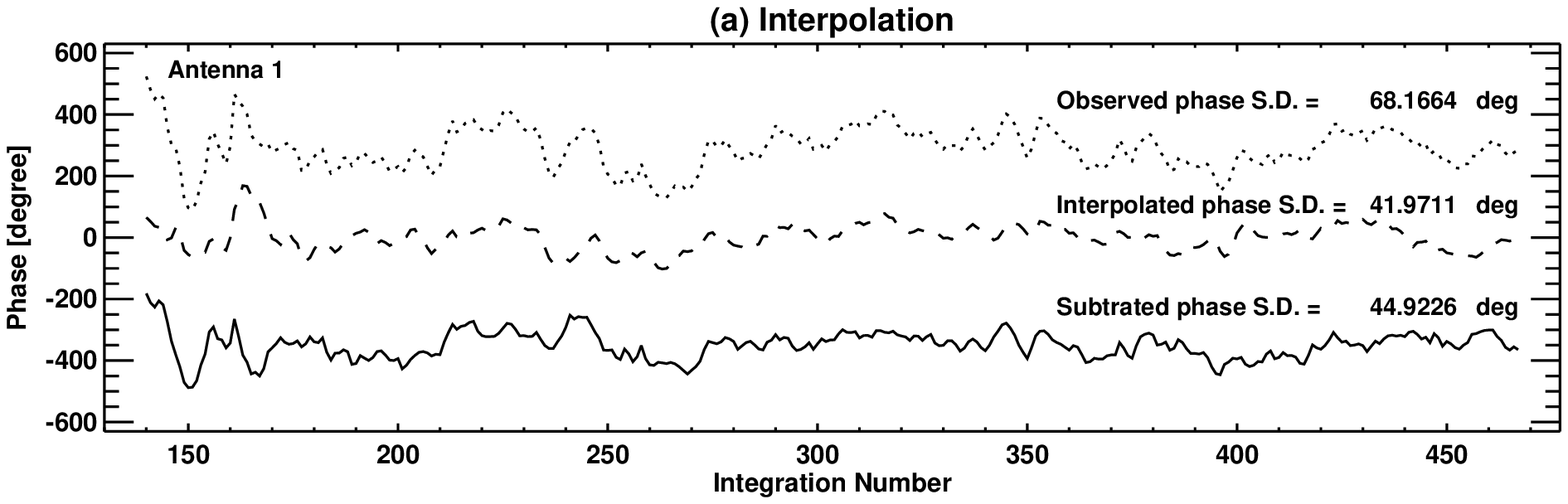}
    \FigureFile(80mm,100mm){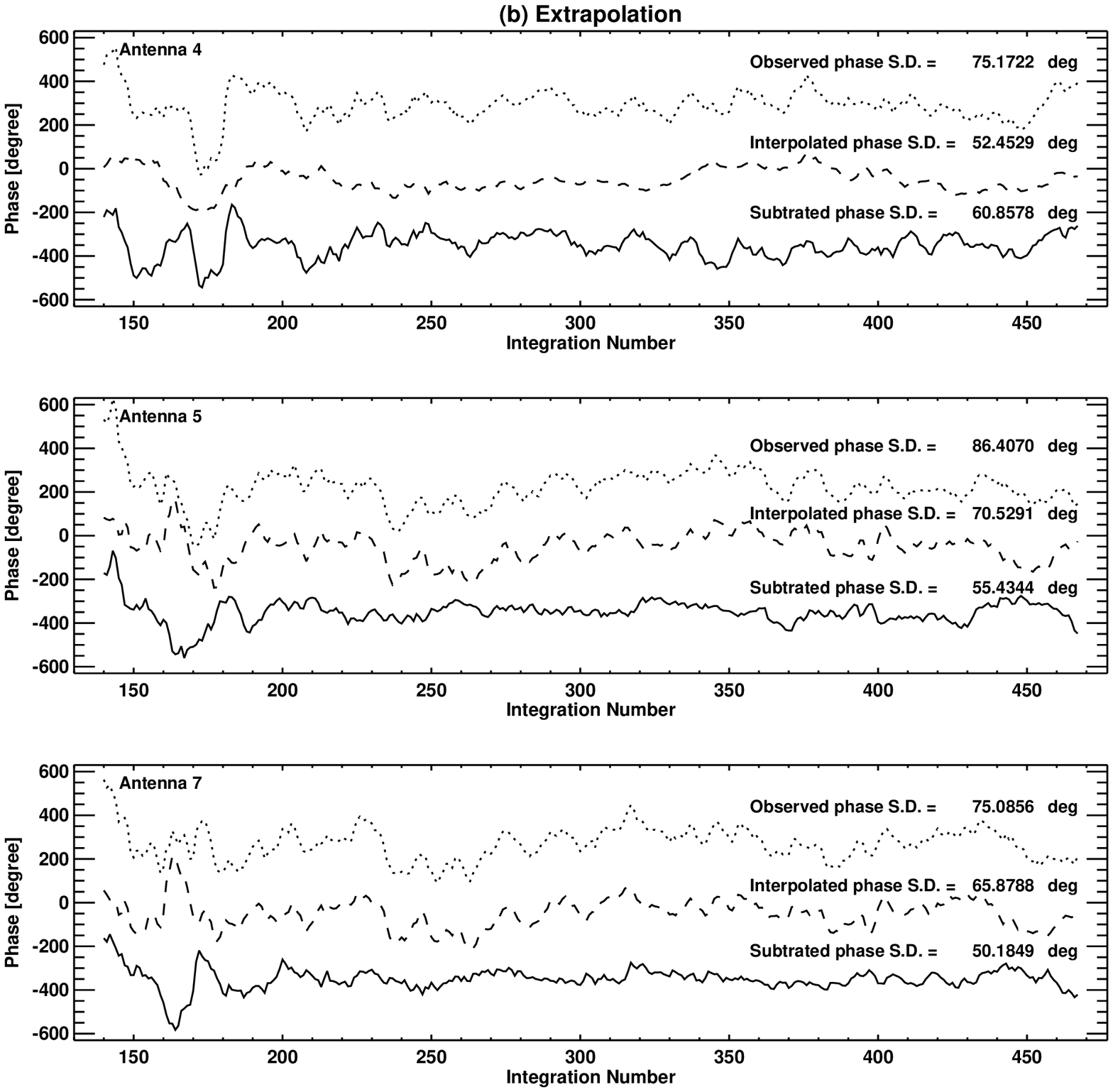}
  \end{center}
  \caption{The time series plots of various phases from the dataset
    040907 between the integration number 140 -- 467 with the
    reference triangle [2, 3, 6].
    Other information is the same as in figure~\ref{fig-resid1}.
    (a) The interpolation results for the antenna 1.
    (b) The extrapolation results for the antennas 4, 5, and 7.}
  \label{fig-resid2}
\end{figure*}

First, we show the examples of the phase fluctuation plots of the
observed data after the antenna-base gain calibration, and of the
interpolated or extrapolated data estimated from a reference triangle
for each interpolated or extrapolated antenna in
figures~\ref{fig-resid1} and \ref{fig-resid2}.
Figure~\ref{fig-resid1} is the plots for the dataset 040826 with the
antenna-base gain calibration referring the antenna 2, and with the
reference triangle of [2, 3, 6] as an example.
Plots in the left and right columns display the results of the
interpolated and the extrapolated antennas, respectively.
Figure~\ref{fig-resid2} is the same plots, but for the dataset
040907.
We also overplotted the subtracted (residual) phase fluctuation plots
in the same figures, which are calculated as follows:
\begin{eqnarray}
  \lefteqn{[{\rm subtracted~(residual)~phase}]} \nonumber \\
    & = & [{\rm observed~antenna~base~gain~calibrated~phase}]
                                                     \nonumber \\
    &   & -~[{\rm interpolated~or~extrapolated~phase}].
  \label{eq-res}
\end{eqnarray}
The subtracted phase tells us how different between the observed and
our estimated interpolated or extrapolated phases are, namely how
effective our phase correction is.

To evaluate the effectiveness of the phase correction quantitatively,
we calculate the standard deviation of our observed,
interpolated/extrapolated, and subtracted phase fluctuations, and
shown in each plot of figures~\ref{fig-resid1} and \ref{fig-resid2}.
It appears that the interpolation scheme leads to smaller residual
phases, while the extrapolation scheme does not always improve the
phase fluctuations.

\begin{figure*}
  \begin{center}
    \FigureFile(160mm,100mm){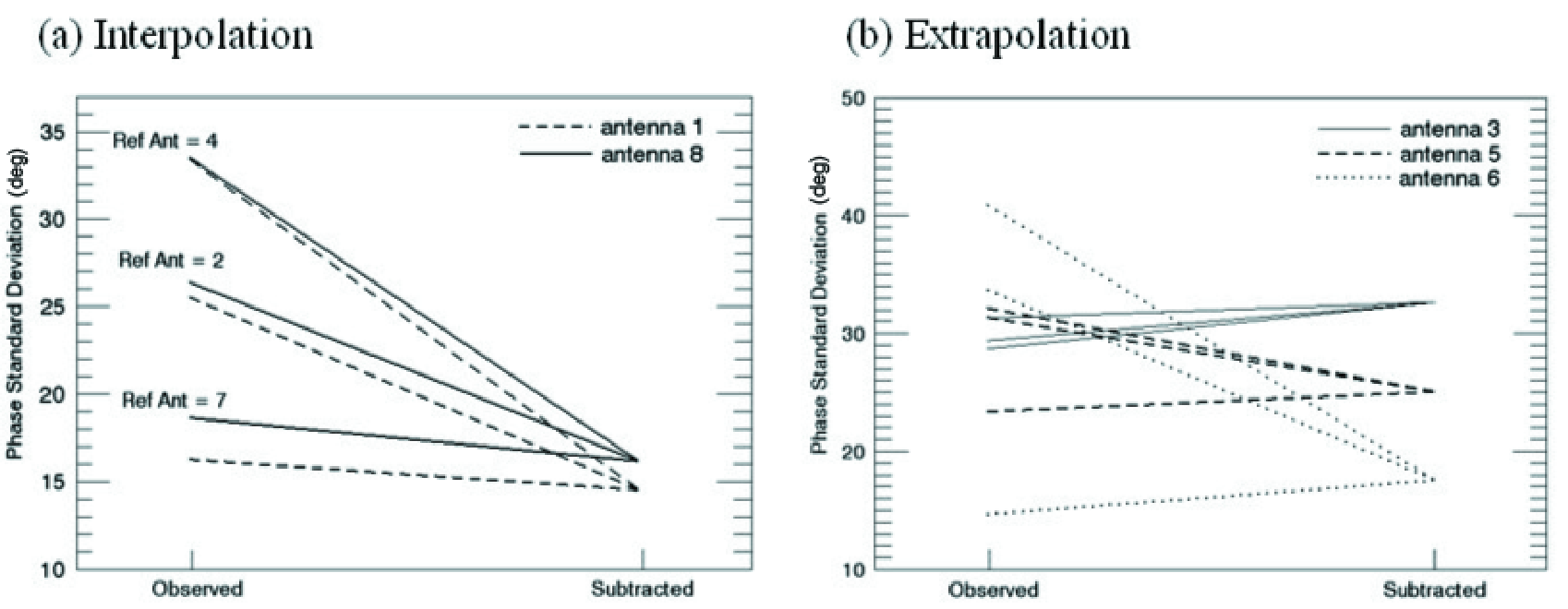}
  \end{center}
  \caption{Observed and subtracted phase standard deviation under the
    same reference triangle [2, 4, 7] and the same integration number
    range 2800 -- 3128, but different reference antennas for the
    antenna-based gain calibration.
    Data calibrated with different reference antennas have different
    observed phase standard deviation, but they all have identical
    phase standard deviations after the phase subtraction.
    (a) The interpolation results for antennas 1 and 8.
    (b) The extrapolation results for antennas 3, 5, and 6.}
  \label{fig-ba}
\end{figure*}

It is worth to note that the observed phase fluctuation of the
dataset 040826 is more stable than that of the dataset 040907, and
the phase fluctuation decreased in both cases (for the interpolation
results).
In addition, even within one dataset, the degree of phase fluctuation
and that of phase correction change drastically.
For example, in the dataset 040826, the phase fluctuation of the
subtracted phase changes a lot between the integration number ranges
of 2321 -- 2799 and 2800 -- 3128; the standard deviation of the phase
fluctuation improved a bit for the former case, but improved a lot
for the latter case (figure~\ref{fig-resid1}).
This suggests that under whatever weather condition (as far as there
is no $2\pi$ ambiguity; see the next paragraph), the interpolation
phase correction works.

When the phase fluctuation is too large, this phase correction does
not work effectively in either interpolation or extrapolation.
This effect is obvious in the integration number range of 140 -- 199
of the 040907 dataset.
We find out that the main reason of this failure is due to the $2\pi$
ambiguity of the observed phase.
Interferometers can measure the phase only within $\pm\pi$, so if the
phase fluctuates largely, the observed phase wraps within $\pm\pi$,
and it is difficult to recover the actual phase fluctuations larger
than $\pm\pi$.

Hereafter, we separate each dataset into two integration number
ranges, and these are shown in table~\ref{tab-def}.
We divided both of our two datasets into earlier and later parts.
The foreparts of 040826 and 040907 are the integration number ranges
of 2321 -- 2799 and 140 -- 199, respectively, which have larger
variations on phase than the later parts of 2800 -- 3128 and 200 --
467.

\begin{table}
  \begin{center}
  \caption{Two datasets divided into two parts based on their
    observed phase variations.}
  \label{tab-def}
  \begin{tabular}{ccc}
    \hline
                 & \multicolumn{2}{c}{Integration Number Range} \\
    \cline{2-3}
                 &    040826    &   040907   \\
    \hline
    Earlier Part & 2321 -- 2799 & 140 -- 199 \\
    Later Part   & 2800 -- 3128 & 200 -- 467 \\
    \hline
  \end{tabular}
  \end{center}
\end{table}

\subsection{Re-Define the Phase:
  Phase Refers to the Center of the Reference Triangle}
\label{sect-res-redef}

The degree of the improvement of the phase fluctuation after the
phase correction, however, depends on the reference antenna of the
antenna-based gain calibration.
If the interpolated antenna is close to the reference antenna, the
improvement of the phase fluctuation is small, but if the
interpolated antenna is far from the reference antenna, the
improvement of the phase fluctuation is large.
In addition, the final results, namely the residual phase
fluctuations, do not depend on reference antennas.
In figure~\ref{fig-ba}, we show examples of the observed and
subtracted phases with different reference antennas for the
antenna-based gain calibration under the same reference triangle
[2, 4, 7].
As can be seen, the standard deviations of the observed phases
largely depend on the reference antennas, but that of the subtracted
phases converges into one.
Furthermore, in case of the interpolation scheme
(figure~\ref{fig-ba}a), subtracted phases improve in all the three
reference antenna cases, but in case of the extrapolation scheme
(figure~\ref{fig-ba}b), subtracted phases improve in some reference
antenna cases and some do not.

\begin{figure*}
  \begin{center}
    \FigureFile(160mm,100mm){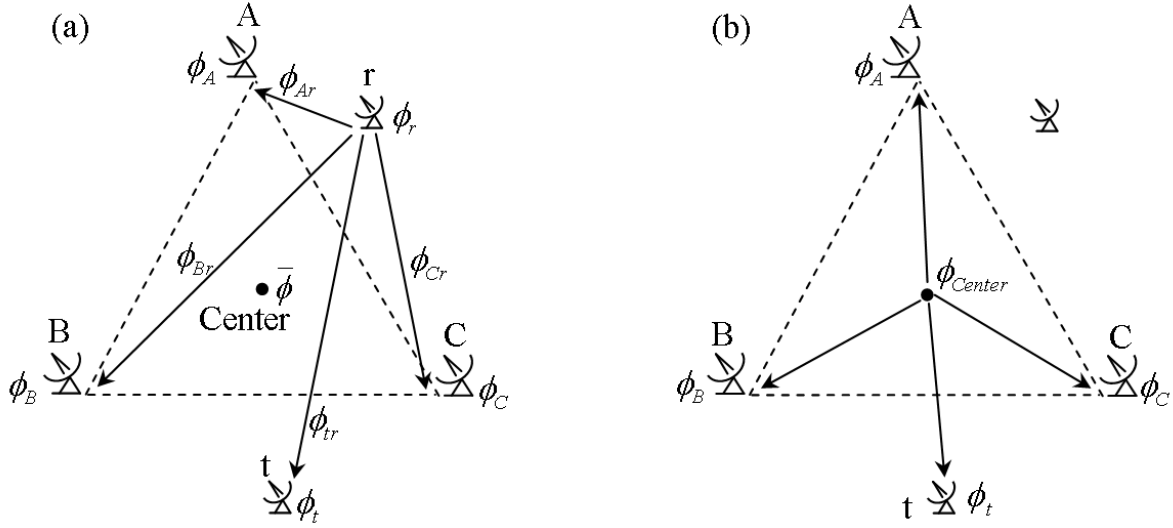}
  \end{center}
  \caption{The simple diagrams of the phase conversion from (a) the
    observed phase to (b) the phase measured from the center of the
    reference triangle.
    The three antennas A, B, and C are located at the corners of the
    reference triangle.
    The reference antenna for the antenna-base gain calibration, $r$,
    and the target antenna whose phase is to be corrected, $t$, are
    also shown in the plots.}
  \label{fig-convert}
\end{figure*}

These can be explained as follows:
Observed phase with the antenna X using the antenna $r$ as the
reference antenna for the antenna-based gain calibration,
$\phi_{Xr}$, can be expressed as
\begin{equation}
  \phi_{Xr} = \phi_{X} - \phi_{r},
  \label{eq-ar}
\end{equation}
where $\phi_{X}$ and $\phi_{r}$ are the absolute phase observed with
the antenna X and the reference antenna $r$.
The estimated phase with interpolation or extrapolation,
$\phi'_{Xr}$, can be expressed as
\begin{equation}
  \phi'_{Xr} = \phi_{X, estimated} - \phi_{r},
  \label{eq-ardash}
\end{equation}
where $\phi_{X, estimated}$ is the absolute estimated phase at the
antenna X.
Both of the above equations depend on the absolute phase of the
reference antenna $r$.
On the other hand, the final result, which is the subtraction between
the observed and the extimated phases, is expressed as
\begin{eqnarray}
  \phi_{Xr} - \phi'_{Xr} & = & \phi_{X} - \phi_{r}
                               - (\phi_{X, estimated} - \phi_{r})
                                                         \nonumber \\
                         & = & \phi_{X} - \phi_{X, estimated},
  \label{eq-ar-ardash}
\end{eqnarray}
which does not depends on the reference antenna, and therefore the
subtracted phase converged into one result, as shown in
figure~\ref{fig-ba}.

To evaluate the effectiveness of the phase correction more
quantitatively, we re-define the phase to that refers to the center
of the reference triangle.
Consider a reference triangle composed of three antennas, A, B, and
C (figure~\ref{fig-convert}a).
The observed phases for these three antennas after the antenna-base
gain calibration using the reference antenna $r$, $\phi_{Ar}$,
$\phi_{Br}$, and $\phi_{Cr}$, can be expressed as
\begin{equation}
  \phi_{Ar}=\phi_A-\phi_r, \\
  \label{eq-phiar}
\end{equation}
\begin{equation}
  \phi_{Br}=\phi_B-\phi_r, \\
  \label{phibr}
\end{equation}
\begin{equation}
  \phi_{Cr}=\phi_C-\phi_r,
  \label{eq-phicr}
\end{equation}
where $\phi_{A}$, $\phi_{B}$, $\phi_{C}$, and $\phi_{r}$ are the
absolute phases at the three antennas of the reference triangle and
at the reference antenna $r$, respectively.
We then consider the phase for the interpolated or extrapolated
antenna $t$.
The absolute and the observed phases for the antenna $t$ can be
expressed as $\phi_{t}$ and $\phi_{tr}$, respectively, and the
relation between $\phi_{t}$ and $\phi_{tr}$ can be written as
\begin{equation}
  \phi_{tr}=\phi_t-\phi_r.
  \label{eq-phitr}
\end{equation}

Now, we re-define the phase, which is, not measure from the reference
antenna $r$, but from the center of the triangle (see
figure~\ref{fig-convert}b).
First, we define $\overline{\phi}$ as
\begin{eqnarray}
  \overline{\phi} & \equiv & \frac{1}{3}(\phi_{Ar}+\phi_{Br}+\phi_{Cr})
                                                               \nonumber \\
                  & = & \frac{1}{3}(\phi_{A}+\phi_{B}+\phi_{C}) - \phi_{r}.
  \label{eq-avephi}
\end{eqnarray}
We can consider this phase as the phase of the center of the triangle
relative to the reference antenna $r$.
If we subtract the phase $\overline{\phi}$ from other phases, these
phases will be the phases refer to the center of the triangle:
For example, the phase of antenna $t$ relative to the center of the
reference triangle can be expressed as
\begin{equation}
  \phi_{tr} - \overline{\phi} = \phi_t - \phi_r - \overline{\phi}.
  \label{eq-phitr-avephi}
\end{equation}
If we substitute $\overline{\phi}$ in equation~(\ref{eq-avephi}) into
this equation, the equation can be written as
\begin{eqnarray}
  \phi_{tr} - \overline{\phi} & = & \phi_t - \phi_r
    - [\frac{1}{3}(\phi_{A} + \phi_{B} + \phi_{C}) - \phi_r ]
                                                     \nonumber \\
  & = & \phi_t - \frac{1}{3}(\phi_{A} + \phi_{B} + \phi_{C}).
\end{eqnarray}
The final form of this equation does not include the phase of the
reference antenna $\phi_r$, and only depends on the absolute phase of
the target antenna $\phi_t$ relative to the absolute phase of the
center of the triangle, $(\phi_{A}+\phi_{B}+\phi_{C})/3$.

\begin{figure*}
  \begin{center}
    \FigureFile(160mm,100mm){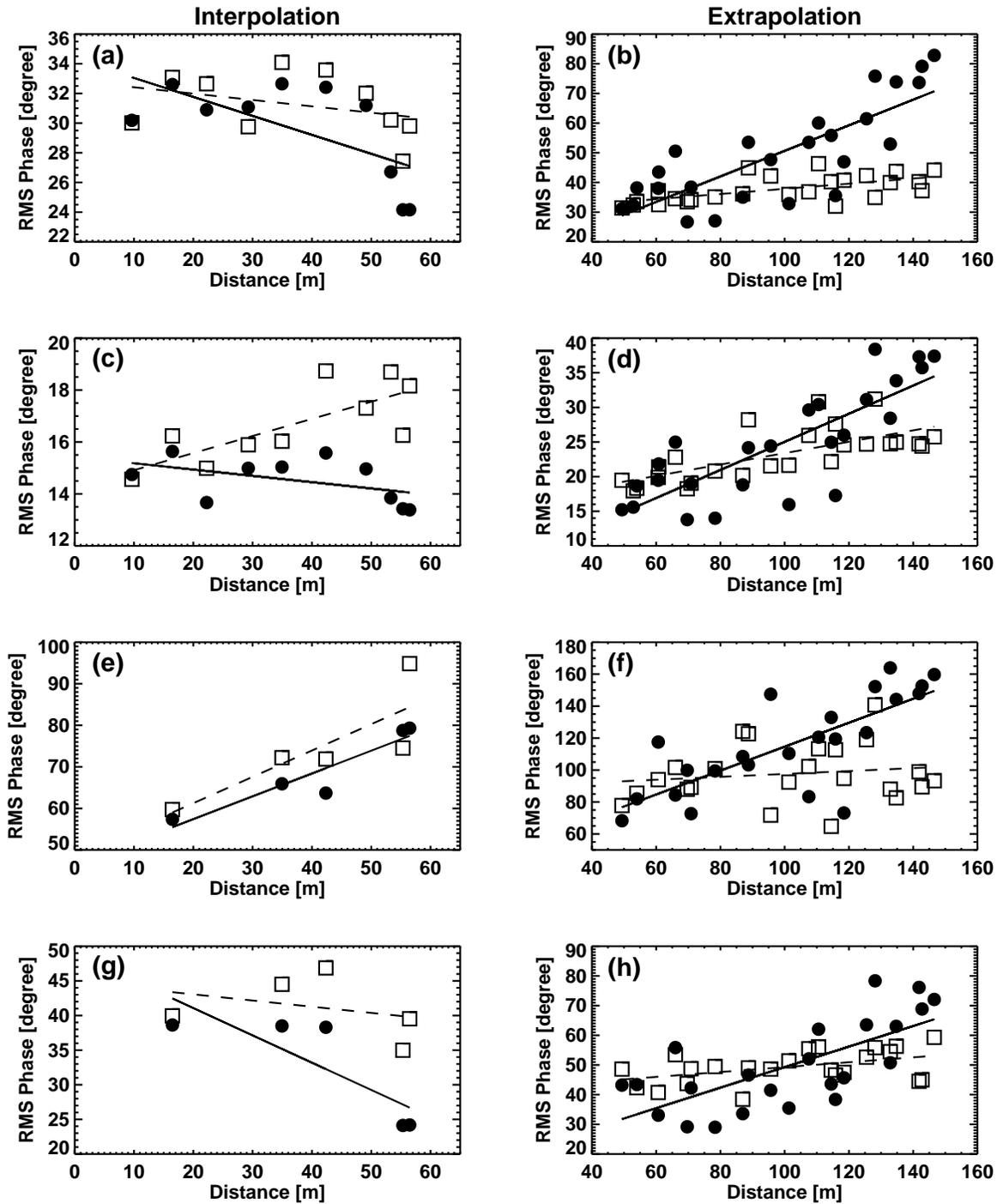}
  \end{center}
  \caption{RMS phase plots of the uncorrected (observed; open
    squares) and corrected (subtracted; filled circles) phases for
    the interpolated and extrapolated antennas.
    Dashed and solid lines are the linear fittings of the uncorrected
    and corrected phases, respectively.
    Plots in the left column are the interpolated results, and those
    in the right column are the extrapolated results.
    (a) Interpolated results for the integration number range 2321 --
        2799 of the dataset 040826.
    (b) Extrapolated results for the integration number range 2321 --
        2799 of the dataset 040826.
    (c) Interpolated results for the integration number range 2800 --
        3128 of the dataset 040826.
    (d) Extrapolated results for the integration number range 2800 --
        3128 of the dataset 040826.
    (e) Interpolated results for the integration number range 140 --
        199 of the dataset 040907.
    (f) Extrapolated results for the integration number range 140 --
        199 of the dataset 040907.
    (g) Interpolated results for the integration number range 200 --
        467 of the dataset 040907.
    (h) Extrapolated results for the integration number range 200 --
        467 of the  dataset 040907.
  }\label{fig-sd}
\end{figure*}

\subsection{Comparison between Actual and Subtracted Phase
  Fluctuations}
\label{sect-res-comp2}

We then compare the relationship between the distance from the
interpolated/extrapolated antennas to the center of the reference
triangle and the subtracted/unsubtracted (corrected/uncorrected)
root mean square (rms) of the phase fluctuation, which is depicted in
figure~\ref{fig-sd}.
We separated the interpolated and the extrapolated antennas, and also
separated the large and small phase fluctuation data for each dataset
(see table~\ref{tab-def}).
the uncorrected (observed) phase fluctuations are similar or increase
with the distance from the center of the reference triangle.
The corrected (subtracted) phase, on the other hand, shows different
behavior between the interpolated and extrapolated results.

The interpolated results show improvement in phase fluctuation,
especially for the longer distance antennas, and in most cases, rms
phase turned to be constant, or sometimes better in the longer
distance antennas than the shorter distance antennas.
This suggests that the phase correction efficiency depends mainly on
the distance; better phase correction efficiency at longer distance.
This may be because the longer distance antennas are closer to an
edge of a reference triangle, namely close to a baseline of two
reference antennas (i.e., one dimension), and the estimated phase
turns to be closer to the real phase than the two dimension case (i.e.,
estimating phase close to the center of the reference triangle with
three antennas).

Exception for this result is the earlier part (the integration number
range of 140 -- 199) of the dataset 040907 (figure~\ref{fig-sd}e; see
also figure~\ref{fig-resid2}a), which is largely affected by the
$2\pi$ ambiguity.
This interpolation result suggests that the phase correction scheme
for the ACA will work properly, as far as the phase does not
fluctuate too large ($<2\pi$).

The extrapolated results, on the other hand, show improvement in some
cases, but often turn to be worse.
Generally, the phase fluctuation increases with the distance from the
center of the reference triangle.
This indicates that the extrapolation of the phase screen does not
work well for the phase correction.
There are huge rises of the corrected phases around the distance of
140~m from the center of the reference triangle in all the
extrapolation results (see figures~\ref{fig-sd}b, d, f, and h;
compare with the observed rms phases, the corrected rms phases
increase significantly).
These data points are mostly the extrapolated phases of the antennas
2, 3, and 5 from the reference triangle [2, 4, 6], [2, 4, 7],
[3, 4, 6], and [4, 5, 6].
We will discuss this later in section~\ref{sect-dis-140m}.

\subsection{Comparisons of RMS Phase with Temporal Structure Function}

\begin{figure}
  \begin{center}
    \FigureFile(80mm,100mm){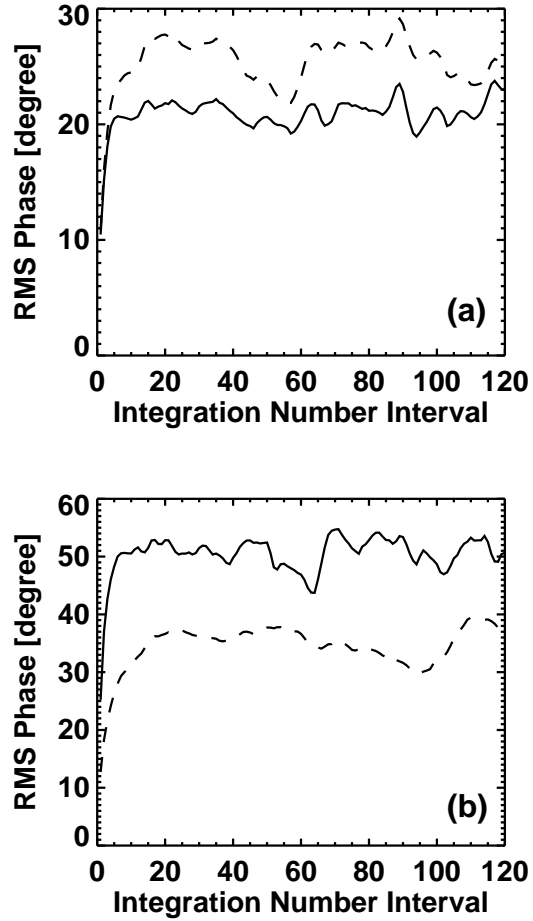}
  \end{center}
  \caption{Examples of the temporal structure function plots for the
    corrected (solid lines) and uncorrected (dashed) rms phases
    with the reference triangle [2, 4, 6] from the dataset 040826.
    (a) The interpolated antenna 1 temporal structure function plots
        for the integration number range 2800 -- 3128.
    (b) The extrapolated antenna 3 temporal structure function plots
        for the integration number range 2800 -- 3128.}
  \label{fig-temp}
\end{figure}

To characterize the troposheric fluctuation, the structure function
\citep{tat61} is often used.
Here, to evaluate the time variation of phase quantitatively, we use
the temporal structure function, $D_{\phi}(\tau)$, which can be
defined as
\begin{equation}
  D_{\phi}(\tau) \equiv \langle[\Phi(t+\tau)-\Phi(t)]^{2}\rangle,
\label{eq-tsf}
\end{equation}
where $\tau$ is the characteristic integration time interval
(in our data, this corresponds to the integration number
interval),
and $\Phi(t)$ is the phase at integration time $t$.
The angle bracket ``$\langle\rangle$'' means the time ensemble.
We denote $\phi_{\rm{rms}}$ as the rms of the temporal structure
function $\sqrt{D_{\phi}}$.
We compare $\phi_{\rm{rms}}$ of the corrected and uncorrected phase
for the interpolated and extrapolated data as a function of
integration time intervals.
We show some examples in figure~\ref{fig-temp}, which displays the
temporal structure functions before and after the interpolation
scheme on the antenna 1 and that in the extrapolation scheme on the
antenna 3 with the reference triangle [2, 4, 6].
The plots show that rms phase (corresponds to $\phi_{\rm{rms}}$
above) rises to a maximum value as integration number interval
(corresponds to $\tau$ in the equation~\ref{eq-tsf}) increases, and
tend to be flat at this maximum value.
If the maximum value for the corrected phase turns to be lower than
the uncorrected phase, the temporal structure function plots tell
that the phase correction worked well.
The interpolation scheme improves the phase, while the extrapolation
scheme makes the phase fluctuation worse.

\section{Discussion}
\label{sect-dis}

In the previous section, we showed that the interpolation phase
correction scheme worked well in our experiments, which supports the
use of this scheme for the ACA.
Here we discuss the possible explanations for the success of the
interpolation scheme and failure of the extrapolation scheme,
the application of our results to the ACA phase correction scheme,
and the validity of the ``flozen-flow'' model.

\subsection{Interpolation and Extrapolation Phase Correction Schemes}
\label{sect-dis-corr}

\begin{figure}
  \begin{center}
    \FigureFile(80mm,100mm){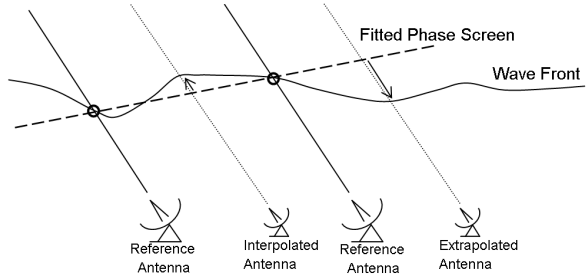}
  \end{center}
  \caption{The wavefront corrected by the fitted phase screen.
    Under this one-dimensional example, the interpolation estimations
    have two reference antennas being ``two-side'' boundary
    condition, while the extrapolation estimations have only
    ``one-side'' boundary condition.}
  \label{fig-1d}
\end{figure}

\begin{figure}
  \begin{center}
    \FigureFile(80mm,50mm){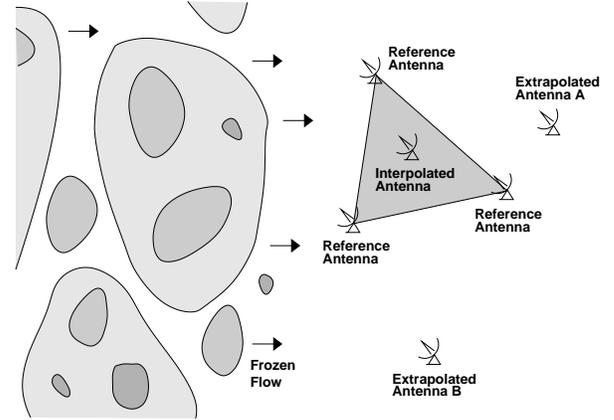}
  \end{center}
  \caption{Schematic diagram showing the antenna configurations and
    the frozen flow of water vapor clumps.}
  \label{fig-frozen}
\end{figure}

The results of our experiments exhibit that the interpolation scheme
provides a better phase correction than the extrapolation approach.
The difference between these two schemes can be compared in
figure~\ref{fig-1d} (simplified to one-dimensional example).
The interpolation scheme is calculated under the three boundary
conditions of the three antenna phases surrounding the antennas to be
corrected, while the extrapolation scheme is calculated for antennas
outside a reference triangle, namely only one boundary condition on
one side, and no boundary condition on the other sides.
Therefore, the extrapolation results deviate more than the
interpolation results due to more degrees of freedom or more
uncertainties.
Furthermore, the distortion of the wave front is caused by the
variations of the water vapor distribution in the troposphere that
move across an interferometer.
The situation between the atmosphere and the interferometer is
depicted in figure~\ref{fig-frozen}.
Smaller scale water vapor ``clumps'' cause a smaller phase
variations, and larger scale ``clumps'' cause a larger phase
variations.
In case of the interpolation scheme,
the water vapor clumps larger than the separation of the reference
antennas can be corrected by the phase screen, and only the small
fluctuation remains.

On the other hand, since the extrapolation scheme is calculated
with only one boundary condition, the water vapor clumps detected
with the reference triangle may not be related to that detected with
the extrapolated antennas, and therefore resulted as a large
variation of phase (larger the distance from the center of the
reference triangle, larger the phase variations; but see
section~\ref{sect-dis-140m} for the difference in the phase
correction results between the antennas located along or
perpendicular to the wind direction from the reference triangle).

\subsection{Differences between Our Experiments and the ACA Phase
	Correction Scheme}
\label{sect-dis-diff}

Although our results support the ACA phase correction scheme, there
are some differences between our experiments and the proposed scheme.
One is the difference in sites; our experiments were done at the
summit of Mauna Kea, Hawaii, but the ALMA site is located at
Chajnantor, Chile.
Therefore the atmospheric conditions between these two sites may
have different characteristics.
However, past several site testing results using radio seeing
monitors indicate that the structure functions, which characterizes
the water vapor clumps in the atmosphere, of these two sites do not
differ much \citep{hol95b}.
Hence this point will not be a problem.

The other is the difference in the antenna configuration of our tests
and the actual configuration of the WVRs in the ACA, namely the
configuration of a phase screen.
Our experiments uses three antennas to create a phase screen, and
therefore one phase plane is naturally created without any offset
from measured phases.
The ACA, however, uses four WVRs, so that a fitting is needed to
create a phase screen, which produces some difference between the
measured phases and the estimated phase screen.
This difference may create some errors in the phase correction, which
leads to larger phase fluctuations after the correction than our
results.
Indeed, as mentioned in section~\ref{sect-res-comp2}, the phase
correction works better near the edge of the reference triangle
(i.e., near one baseline), supporting this concern.
It would be the future study to compare the phase correction results
using phase screens derived from three antennas and that from four
antennas.

\subsection{Extraordinary 140~m Phase Fluctuations}
\label{sect-dis-140m}

In general cases, the phase fluctuation gradually increases with the
increase of distance from the center of the reference triangle to the
interpolated or extrapolated antennas in our experiments.
However, as mentioned in section~\ref{sect-res-comp2}, the corrected
phases of the extrapolation scheme suddenly rise up around the
distance from the center of the reference triangle of 140~m, much
more (almost twice worse) than the original observed phase
fluctuation.

\begin{figure}
  \begin{center}
    \FigureFile(80mm,50mm){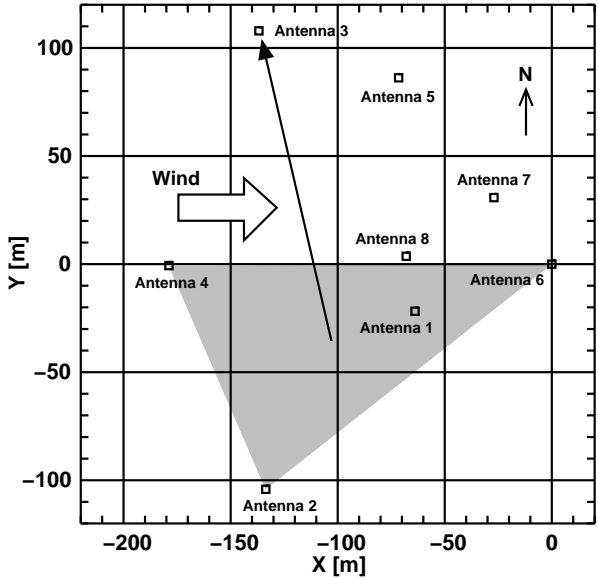}
  \end{center}
  \caption{Schematic diagrams of phase screens created by reference
    antennas and the distance from the center of the reference
    triangle to the extrapolated antennas.
    The black arrow shows an example of the extrapolated antenna 3
    with the distance from the center of the reference triangle
    [2, 4, 6] of around 140~m.
    The white arrow is the rough direction of the prevailing wind
    of an observed day.}
  \label{fig-140m}
\end{figure}

The large scatter is possibly due to the orientation (direction) of
the center of the reference triangle to the extrapolated antennas.
Figure~\ref{fig-140m} presents an example configuration of the
reference triangle [2, 4, 6] for the extraordinarily large rms phase
fluctuation data point for the antenna 3.
The extraordinary 140~m phase fluctuation data all have similar
orientation, which is almost along the north-south direction.
The meteorological parameters on the two observation days show that
the prevailing wind direction is either east or west.
This wind direction is almost perpendicular to the orientations of
the antenna configurations with the extraordinary 140~m fluctuations
on both days (figure~\ref{fig-140m}).

A possible explanation for this extraordinary phenomenon is as
follows:
The time variations of the atmospheric phase are usually approximated
by a ``frozen-flow'' model \citep{tay38,dra79}:
The time scale to develop/cease a turbulence is much longer than
the time taken for a turbulent field to pass across the reference
triangle by wind.
Hence the turbulence of water vapor is generally `frozen' in the
atmosphere, and the wind transport the water vapor turbulences above
an interferometer without changing the size of the turbulences
(figure~\ref{fig-frozen}).
Under this ``frozen-flow'' hypothesis, the antennas outside the
reference triangle but along the wind direction (for example, the
extrapolated antenna A in figure~\ref{fig-frozen}), the observed
phase will not be largely different from the estimated phase by the
extrapolation of the phase screen, since the water vapor turbulences
that pass in front of the antennas are similar to that pass in front
of the reference triangle.
Therefore the phase fluctuation after the phase correction gradually
increases with the distance from the center of the reference
triangle.
Indeed, some of the corrected phases for the extrapolated antennas
are improved from the uncorrected phases (see figure~\ref{fig-sd}),
and most of those extrapolated antennas are located along the wind
direction from the reference triangles.
However, the antennas outside the reference triangle but
perpendicular to the wind direction (for example, the extrapolated
antenna B in figure~\ref{fig-frozen}), the observed phase will be
largely different from the estimated phase by the extrapolation of
the phase screen, since the water vapor turbulences that pass in
front of the antennas are different from that pass in front of the
reference triangle.
Therefore the phase fluctuation after the phase correction generally
make results worse.

The location between the antennas that caused the ``140~m phase
fluctuation'' and the center of the reference triangle is almost
perpendicular to the wind direction of the observed days, and other
antennas are along the wind direction.
We therefore conclude that the results of the extrapolation scheme
including the ``140~m phase fluctuation'' can be explained by
``flozen-flow'' model.

\section{Conclusions}

We performed an interferometric phase correction with the
interpolation or extrapolation of the phase screen defined by
three reference antennas using the SMA.
This interpolation method is proposed for the ACA in the ALMA.

According to the comparisons of the standard deviations of corrected
and uncorrected phases, relations between the phase standard
deviation and the distance from the center of the reference triangle,
and the temporal structure functions of the rms phase, the
interpolation scheme improves phase fluctuation while the
extrapolation scheme does not.
This result can be explained by the boundary conditions of phase in
these schemes; in case of the interpolation scheme, the phase
corrected antenna is inside the triangle of three reference antennas,
so the phase inside the triangle can be well defined (more known
boundary conditions, more precisions or less phase errors and
deviations).
The extrapolation scheme, on the other hand, only has partial
boundary conditions, and therefore less precision.

In the extrapolation scheme results, there is a sudden large phase
fluctuation at the distance from the center of the reference triangle
of about 140~m.
According to the meteorological parameters on those observing dates
and the antenna configurations, this ``140~m phase fluctuation'' is
occurring only at the antennas located from the center of the
reference triangle perpendicular to the wind direction.
This ``140~m phase fluctuation'' can be explained by the frozen-flow
model.

Although there are some differences between the configuration of our
experiments and that in the proposed phase correction scheme for the
ACA, our results based on the actual observations and the simulation
results done by \citet{asa05} promise the success of the phase
correction for the ACA.

\bigskip

We would like to thank Kazushi Sakamoto and Koh-Ichiro Morita for
fruitful discussion.
The Submillimeter Array is a joint project between the Smithsonian
Astrophysical Observatory and the Academia Sinica Institute of
Astronomy and Astrophysics and is funded by the Smithsonian
Institution and the Academia Sinica.

\end{document}